\documentclass[11pt,twoside]{article}

\usepackage{asp2004}
\usepackage{epsf}
\usepackage{psfig}
\usepackage{lscape}

\begin{document}
\title{Very Cool Close Binaries}   
\author{J. Scott Shaw}   
\affil{Department of Physics and Astronomy, University of Georgia, Athens GA, 30602, USA}    
\author{Mercedes L\'{o}pez-Morales}   
\affil{Carnegie Institution of Washington, Department of Terrestrial Magnetism,
5241 Broad Branch Rd. NW, Washington DC 20015, USA
}    

\begin{abstract} 
We present new observations of cool $<$6000K and low mass $<$1$M_{\sun}$ binary systems that have been discovered by searching several modern stellar photometric databases. The search has led to a factor of 10 increase in the number of known cool close eclipsing binary systems.
\end{abstract}


\section{Introduction}   
For a long time the lower main sequence was neglected due to the difficulty in modeling stars of such low mass and due to scarcity of observations of cool close eclipsing binaries. Recently there has been much progress on the side of theory with models of, among others, Siess et al. (1997), Baraffe et al. (1998), and Yi, S. et al. (2001).   However, the observational efforts have lagged behind.  Our knowledge of the fundamental parameters stellar masses and radii comes primarily from stars that are eclipsing and double lined spectroscopic binaries.  Before 2003 only three systems with M-type components were well studied: YY Gem (Torres \& Ribas 2002, Leung \& Schneider 1978), CM Dra   (Metcalfe et al. 1996) and CU Cnc (Delfosse et. al 1999; Ribas 2003). The low number of known binary systems with cool $<$6000K and low mass $<$1$M_{\sun}$ has hindered our understanding of the lower main sequence. In the last three years the number of well know systems has tripled to include: OGLE BW5 V038 (Maceroni  \& Montalb‡n 2004), GU Boo   (L\'{o}pez-Morales \& Ribas 2005), TrES-Her0-07621  (Creevy et al. 2005), EMTSS-6 [RX J0239.1-1028] (Torres et al, 2006), 2MASS J05162881+2607387 (Bayless  \& Orosz, priv. comm), NSVS01031772 (L\'{o}pez-Morales et al., in prep.).  The importance of low mass stars in our understanding of stellar evolution and the contents of the Milky Way is pointed out in our companion paper (L\'{o}pez-Morales \& Shaw, these proceedings). There we also note the discovery of the challenge posed by the new data to our understanding of the theory of stars in the lower main sequence.
\section{The Search for Low Mass Eclipsing Binaries}
Only in the last few years have sky surveys looking for stellar variability attained the faint magnitudes, and number of targets necessary to reveal close eclipsing binaries in the lower main sequence. To date we have searched five recent databases containing photometry of a large number of stars: ROTSE-I (Akerlof et al. 2000), OGLE (Udalski et al. 1992), NSVS (Wozniak et al. 2004), EXPLORE (Mall\'en-Ornelas et al. 2003), and ASAS Ð III (Pojmansky 2002). Listed in Table 1 are (1) the name of the survey, (2) an approximate limiting magnitude for the photometry and its approximate color, (3) the number of stars in the survey in thousands, (4) the number of variable stars found of all types, (5) the number of eclipsing binaries, and (6) the number of cool eclipsing binaries. The four variable systems from OGLE-I were first noted by Maceroni \& Rucinski (1990).

\begin{table}[!ht]
\caption{Surveys Searched for Cool Binaries}
\smallskip
\begin{center}
{\small
\begin{tabular}{llrrrr}
\tableline
\noalign{\smallskip}
Survey&Lim Mag &Stars(K) &Variables&EB&Cool \\
\noalign{\smallskip}
\tableline
\noalign{\smallskip}
ASAS-III&I $<$13&13,000 &$~$150,000&$~$50,000&1\\
NSVS&V=10-15&17,000&$~$16,000&$~$4,000&12\\
OGLE-I&I=14-18&5,000&2,861&1,650&4\\
OGLE-II&I= 11-18&10,000 &221,801&$~$6,000&20\\
EXPLORE&V$<$20&98&143&101&4\\
\noalign{\smallskip}
\tableline
\end{tabular}
}
\end{center}
\end{table}

The results of our search are, so far, 41 new detached eclipsing binaries with masses below 1$M_{\sun}$. The search of the NSVS survey is approximately 50\%  complete. In EXPLORE we have only searched part of the database at southern latitudes.

	To identify the candidates the photometric data was searched for periodicities using two distinct methods: the string/rope length method based on the Lafler-Kinman statistic (Clarke 2000) and the analysis of variance method (Schwarzenberg-Czerney 1989).  Light curves showing periodicity detected by either method were looked at more closely to see if they were cool, detached systems and if their masses were likely to be below 1$M_{\sun}$.  The results are contained in Table 2. The stars are grouped by their survey. The stars in the OGLE-I have names containing BW and MM; the others are self-evident. The columns are (1) name, (2) right ascension (2000), (3) declination (2000), (4) period in days, (5) V magnitude, (6) amplitude of variation, (7) estimated total mass of the system, (8) comments (see below).

\begin{table}[!ht]
\caption{New Low-Mass Eclipsing Binaries}
\smallskip
\begin{center}
{\small
\begin{tabular}{llllclll}
\tableline
\noalign{\smallskip}
Name&RA(2000)&Dec(2000)&P(days)&V&Amp &M1+M2&Comments \\
\noalign{\smallskip}
\tableline
\noalign{\smallskip}

ASAS1647&16 47 55&-08 44 30&0.503264&13.0&0.87&1.3-1.4&LC RV\\

NSVS06507557&01 58 24&+25 21 19&0.51509957&13.1&1.0&0.8-1.0&LC\\
NSVS01828214&02 13 51&+65 46 26&2.21433074&13.2&0.65&$>$1.3&\\
NSVS06848235&05 01 59&+34 47 56&1.5383066&13.4&1.2&1.6-1.7&\\
NSVS07394765&08 25 51&+24 27 05&2.26543128&12.8&1.1&1.7&pLC \\
NSVS02502726&08 44 11&+54 23 48&0.55973867&13.5&1.0&1.1-1.2&LC\\
NSVS07453183&09 16 12&+36 15 32&0.3669689&13.2&0.8&$<$0.6 &LC RV \\
NSVS10441882&13 30 25&+13 49 32&0.5166492&12.8&0.8&1.3-1.4&\\
NSVS01031772&13 45 35&+79 23 48&0.36814246&12.6&1.3&0.8&LC RV S\\
NSVS10653195&16 07 28&+12 13 59&0.56072686&12.6&0.75&1.2&pLC\\
NSVS01178845&17 45 25&+69 18 22&0.4937332&12.2&1.1&$<$1.0  &LC\\
NSVS01286630&18 47 09&+78 42 33&0.38390916&13.3&0.9&1.1&\\
NSVS11868841&23 17 58&+19 17 03&0.60176906&14.2&1.1&1.0&LC\\
BW9 v31&18 00 24.3&-29 51 24.3&1.99655&18.5&0.24&&\\
BW5 v149&18 02 37.4&-30 00 27.2&1.31085&19.7&0.93&&\\
MM7B v101&18 12 02.2&-26 00 31.4&1.20532&19.5&0.53&&\\
BW11 v48&18 01 16.2&-30 17 16.2&3.77077&$>$19.5&0.24&&\\
OGLE432075&17 35 01.6&-27 05 28.5&3.97628808&19.32&0.57&$>$1.6&\\
OGLE431010&17 35 14.5&-27 19 56.3&1.23358142&19.76&0.50&$>$1.3-1.4&\\
OGLE271691&17 48 41.7&-35 12 26.4&0.38797888&19.35&0.45&$>$0.5-0.6&\\
OGLE445255&17 48 55.8&-29 51 12.8&1.59890199&20.88&0.85&0.7&\\
OGLE445095&17 49 05.2&-29 52 04.1&0.77739549&20.26&0.73&0.9-1.0&\\
OGLE441737&17 49 22.7&-30 14 34.5&0.74692839&20.12&0.28&$>$0.85&\\
OGLE051231&17 49 58.0&-30 13 45.6&0.72695040&20.37&1.20&0.5-0.6&\\
OGLE056610&17 50 14.3&-29 33 00.0&2.25488782&19.70&0.60&$>$1.5&\\
OGLE400589&17 50 40.2&-33 34 16.8&0.81628138&19.54&0.50&1.2 &\\
OGLE401621&17 50 43.8&-33 20 18.3&0.46866554&19.96&0.70&0.85&\\
OGLE400678&17 51 03.6&-33 32 51.5&1.62296712&19.50&0.47&$>$1.7&\\
OGLE376098&17 52 22.9&-29 42 05.5&1.52457572&19.90&0.45&$>$1.3&\\
OGLE374040&17 52 39.3&-29 56 37.6&0.46203032&20.03&0.70&0.9-1.0&\\
OGLE252455&17 54 31.1&-32 34 05.3&1.33850920&18.45&0.47&$>$0.8&\\
OGLE393947&17 55 44.7&-29 41 20.3&1.22609925&18.27&0.40&$>$1.7&\\
OGLE344067&17 57 50.4&-29 05 09.8&1.21074128&19.65&0.37&$>$1.3&\\
OGLE011417&18 02 17.9&-30 07 48.5&0.87255347&18.97&2.60&0.8-0.9&\\
OGLE024895&18 04 59.0&-28 29 19.0&0.76093984&20.58&0.50&$>$0.8-0.9&\\
OGLE130381&18 16 49.1&-24 19 20.5&0.79444444&19.51&0.50&$>$0.9&\\
OGLE082082&18 22 56.0&-21 25 25.3&0.61030800&20.60&0.65&$>$1.25&\\
EXPLORE411492&16 25 54.10&-52 38 27.8&1.1024147&&&&\\
EXPLORE720427&16 28 28.01&-52 46 02.0&1.2543001&18.09&0.44&$>$0.95&\\
EXPLORE523625&16 28 52.22&-53 10 55.1&1.5752375&16.94&0.49&$>$1.5&\\
EXPLORE632136&16 29 20.22&-52 56 53.5&0.5735012&17.24&0.52&$>$1.8&\\

\noalign{\smallskip}
\tableline
\end{tabular}
}
\end{center}
\end{table}

We have begun a long-term project to obtain multicolor light curves and radial velocity curves for all of the objects brighter than 14 magnitude.  Our progress is shown in the comments column in Table 2: LC=we have light curves, RV=we have radial velocity curves, pLC=we have partial light curves, S=we have a solution and it is included in L\'{o}pez-Morales \& Shaw (these proceedings) and  L\'{o}pez-Morales et al. (2006, in prep.)

\section{Conclusions}
Recent large stellar photometric surveys are yielding a large number of newly discovered variable stars. Some of these turn out to be the up-to-now rarely known M-type binaries.  We estimate that approximately 1 of 1000 stars surveyed are eclipsing binaries and, of those, about 1 of 1000 are M-type systems.  Our discoveries have helped increase the number of well-studied M-type systems to three times the number known before the year 2003.  Moreover we have three times that number again as many good candidates to study. Future work should insure adequate observations to test stellar models in the lower main sequence.

\acknowledgements We would like to thank the following institutions for providing support for this work: Carnegie Institution of Washington, NASA Astrobiology Institute, Southeastern Association for Research in Astronomy (SARA), University of Georgia at Athens, Instituto de Astrof\'isica de Canarias, National Science Foundation, and the University of North Carolina at Chapel Hill. We also thank our collaborators Jerome A. Orosz (SDSU, USA), Ignasi Ribas (IEEC, Spain), Maria Jes\'{u}s Ar\'{e}valo and Carlos L\'{a}zaro (IAC, Spain), and Guillermo Torres (HarvardÐCfA, USA) for their time and efforts towards this project. We also gratefully acknowledge the hard work of student researchers Yelena Pelimskaya (Lehigh University), Cecelia Hedrick (University of Nebraska), Travis McIntyre (Clemson University) who were supported in part by the National Science Foundation grant AST-0097616 to SARA and of student researchers Susan Chung (University of Georgia) and David Hou (Athens Academy).  M.~L-M. acknowledges research and travel support from the Carnegie Institution of Washington through a Carnegie fellowship.



\begin{thebibliography}{}
\bibitem[Akerlof et al. 2000]{aa2000}
Akerlof, C., Amrose, S., Balsano, R., et al. 2000, AJ, 119, 1901
\bibitem[Baraffe et al 1998]{bar1998}
Baraffe, I., Chabrier, G., Allard, F., \& Hauschildt, P.H. 1998, A\&A, 337, 403
\bibitem[Bayless  \& Orosz 2006]{bo2006}
Bayless, A. \& Orosz J.  2006, private communication
\bibitem[Clarke 2002]{cl2002}
Clarke, D. 2002 A\&A, 386, 763
\bibitem[Creevey et al. 2005]{cr 2005}
Creevey, O. L., Benedict, G. F., Brown, T. R., et al. 2005, ApJ, 625, L127
\bibitem[Delfosse, et al 1999]{del1999}
Delfosse, et. al. 1999 A\&A 341, L63
\bibitem[Leung \& Schneider 1978]{ls1978}
Leung, K-C, \& Schneider 1978 AJ, 83, 618
\bibitem[L\'{o}pez-Morales et al 2006]{los2006}
L\'{o}pez-Morales 2006, private communication
\bibitem[L\'{o}pez-Morales \& Ribas 2005]{lr2005}
L\'{o}pez-Morales, M. \& Ribas, I. 2005, ApJ, 631, 1120
\bibitem[ L'{o}pez-Morales \& Shaw  2006]{ms2006 }
L\'{o}pez-Morales, M \& Shaw, J.S. \&. 2006, these proceedings
\bibitem[Maceroni  \& Montalb\'{a}n 2004]{mm2004}
Maceroni, C  \& Montalb\'{a}n 2004 A\&A 426, 577
\bibitem[Maceroni \& Rucinski 1999]{mr1999}
Maceroni, C. \& Rucinski, S.M. 1999 AJ 118,1819
\bibitem[Mall\'{e}n-Ornelas  et al. 2003]{ma2003}
Mall\'{e}n-Ornelas, G., Seager, S., Yee, H. K. C. et al. 2003, ApJ, 582, 1123
\bibitem[Metcalfe et al 1996]{me1996}
Metcalfe, T. S., Mathieu, R. D., Latham, D. W., \& Torres, G. 1996, ApJ, 456, 356
\bibitem[Pojmanski 2002]{poj2002}
Pojmanski, G. 2002, Acta Astronomica, 52,397
\bibitem[Ribas(2003)]{rib03} Ribas, I. 2003, A\&A, 398, 239
\bibitem[Schwarzenberg-Czerney 1989]{sch1989}
Schwarzenberg-Czerney, A. 1989, MNRAS, 241,153
\bibitem[Siess et al 1997]{sie1997}
Siess, L., Forestini, M., \& Dougados, C. 1997, A\&A, 324, 556
\bibitem[Torres et al 2006]{tlm2006}
Torres G. 2006, private communication
\bibitem[Torres \& Ribas 2002]{tr2002}
Torres, G. \& Ribas, I. 2002, ApJ, 567, 1140
\bibitem[Yi et al. 2001]{yi2001}
Yi, S., Demarque, P., Kim, Y. C. et al. 2001, ApJS, 136, 417
\bibitem[Udalski et al 1992]{ u1992}
Udalski, A., Szymanski, M., Kaluzny, J., Kubiak, M., \& Mateo, M. 1992, AcA 42, 253
\bibitem[Wozniak et al. 2004]{woz2004}
Wozniak, P. R., Vestrand, W. T., Akerlof, C. W., et al. 2004, AJ, 127, 2436
\end{thebibliography}
\end{document}